# Features of incommensurate phases in crystals TlGaSe2 and TlInS2

B.R. Gadjiev

*International University of Nature, Society and Mans "Dubna",*
*141980, 19 Universitetskaya str., Dubna, Moscow Region, Russia*
*gadjiev@uni-dubna.ru*

**Abstract**
The theory of a sequence of phase transitions of high-symmetry-incommensurate-commensurate phase controlled by competing order parameters is investigated. The temperature dependence of dielectric constant is derived. The comparison of the obtained results with experimental data for crystals TlGaSe$_2$ and TlInS$_2$ is discussed.

**Introduction**

In discussion of the incommensurate phases in ideal crystals on the basis of the Landau theory two cases are usually distinguished. In the first of them we have the symmetry of two (or more) component order parameter, which is supposed by the Lifshitz gradient invariant, which leads to spatially modulated phase [1]. In the second case it is absent for one component order parameter [2]. These two cases are called as incommensurate phases of a type II and I respectively. As is known, the one-dimensional modulated incommensurate phases in proper ferroelectrics corresponding to a type II, whereas in improper ferroelectrics — to a type I.

There is a series of works in which the features of static and dynamic properties of incommensurate phases of the types II and I near the normal-incommensurate and the incommensurate-commensurate phase transitions are investigated [1, 2, 3, 4, 5].

In improper ferroelectrics a dielectric constant at the point $T_i$ (the temperature of transition from high-symmetry in an incommensurate phase), does not show any anomalies, whereas in soliton regime incommensurate phase behaves as $\sim (T - T_c)^{-1}$, ($T_c$ is the temperature of transition from incommensurate in a commensurate phase) and the values of dielectric constant coincide in high-symmetry and commensurate phases. In the incommensurate phase the temperature dependence of dielectric constant of proper ferroelectrics is similar to that of dielectric constant of improper ferroelectrics [2, 4, 5]. However, in phase transitions of proper ferroelectrics, dielectric constant in high-symmetry and commensurate phases shows the Curie-Weiss type divergence.

To the present time of physical properties of a number of crystals with an incommensurate phase are enough deeply investigated. The researches of thermodynamic, optical and electrical properties of these phases as a whole have shown the good agreement of theoretical and experimental results [5].

Now there is a question on a character of the phase transition in the layered crystals TlGaSe$_2$ and TlInS$_2$. The experiments on X-ray analysis and the neutrons scattering show, that in high-symmetry phase these crystals have the space group of symmetry $C_{2h}^6$ [6]. The TlGaSe$_2$ and TlInS$_2$ crystals are in the incommensurate phase in a temperature interval from $T_i \approx 120$ K down to $T_c = 110$ K, and from $T_i \approx 220$ K down to $T_c = 200$ K, respectively [7, 8, 9, 10, 11]. A modulation wave vector in both crystals is $\vec{q}_{inc} = (\sigma, 0, 0.25)$, where $\sigma \approx 0.04$ [8, 9, 10,]. The elementary cell of these crystals is quadruplicated in a commensurate phase, that is quite characteristic for improper ferroelectrics [12]. The commensurate phase is polar. However, the symbol of the group of the commensurate phase in these crystals is not determined yet.

On the basis of these data in [12] the theory of the improper ferroelectrics with an incommensurate phase is developed. In particular, the temperature dependence of dielectric constant is obtained in the region of a sequence of phase transitions high symmetry – incommensurate – commensurate phase for crystals of a type TlGaSe$_2$. However, in [9] is shown, that it is not possible to explain the experimental temperature dependence of dielectric constant of compounds TlGaSe$_2$ and TlInS$_2$ by use the results of work [12].

Thus, really, in crystals TlGaSe$_2$ and TlInS$_2$ temperature dependence of dielectric constant in the region of a sequence of phase transitions high symmetry – incommensurate – commensurate phase shows properties, in the certain sense characteristic for proper ferroelectrics with an incommensurate phase [13]. In particular, the measurements have shown, that both the higher $T_i$, and the lower $T_c$, the low-frequency dielectric susceptibility submits to the law Curie-Weiss, which is characteristic for proper ferroelectrics with incommensurate phase [13]. However, in an incommensurate phase of both compounds dielectric constant rather weakly depends on temperature, that is not characteristic neither for proper, nor for improper ferroelectrics [9,13].

In further was shown, that as opposed to structural researches, the Raman optical experiments in compounds TlGaSe$_2$ indicate on a presence of a soft mode with temperature dependence, being characteristic for proper ferroelectrics [14]. In this case there is a not clear reason quadruple of volume of an elementary cell along an axis $z$.

Thus, on the one hand, the structural experiments show, that the crystals TlGaSe$_2$ in a temperature interval from $T_i \approx 120$ K down to $T_c = 110$ K, and TlInS$_2$ from $T_i \approx 220$ K down to $T_c = 200$ K, are in an incommensurate phase [7, 8, 9]. And a modulation wave vector in both crystals is, where $\vec{q}_{inc} = (\sigma,0,0.25)$, where $\sigma \approx 0.04$ [8, 9, 10]. Hence, the phase transition has the character of improper ferroelectrics. On the another hand, the Raman optical experiments show, that in spectra there is only soft mode, and there are no some appreciable attributes of phase transitions with quadruple of volume of an elementary cell [14]. Last, on seen, is caused very small dispersion of fluctuations on a $z$ direction.

Proceeding from above-stated it is possible to conclude, that, really, in these crystals the phase transitions are accompanied by two soft modes corresponding to wave vectors $\vec{\Gamma} = (0,0,0)$ and $\vec{q}_0 = (0,0,\frac{1}{4})$ of the space group $C_{2h}^6$. In the present work the theory of a sequence of phase transitions with an incommensurate phase, controlled by competing order parameters, is investigated.

**Theory**

Let us consider a structure undergoing a sequence of phase transitions of a type high-symmetry – incommensurate – commensurate phase controlled by two major order parameters.

Let one-dimensional order parameter, namely the $P_y$–component of vector polarization, be transformed by irreducible representation $A_u$ of the space group $C_{2h}^6$. The two-dimensional order parameter is transformed by irreducible representation $D^{*\vec{q}_0}$ of the space group $C_{2h}^6$, corresponding to the point $\vec{q}_0 = (0,0,0.25)$. The symmetry operations and irreducible representation corresponding to the wave vector $\vec{q}_0 = (0,0,0.25)$ are given in Table 1.

**Table 1.** Two dimensional irreducible representation $D^{*\vec{q}_0}$ of the space group $C_{2h}^6$, corresponding to the wave vector $\vec{q}_0 = (0,0,0.25)$, where $\varepsilon = \exp(-i\pi/4)$ and $\vec{\tau} = (0,0,c/2)$

| $C_{2h}^6$ | $\{e/0\}$ | $\{\sigma_h/\vec{\tau}\}$ | $\{I/0\}$ | $\{C_2/\vec{\tau}\}$ |
|---|---|---|---|---|
| $D^{*\vec{q}_0}$ | $\begin{pmatrix} 1 & 0 \\ 0 & 1 \end{pmatrix}$ | $\begin{pmatrix} \varepsilon & 0 \\ 0 & \varepsilon^* \end{pmatrix}$ | $\begin{pmatrix} 0 & 1 \\ 1 & 0 \end{pmatrix}$ | $\begin{pmatrix} 0 & \varepsilon \\ \varepsilon^* & 0 \end{pmatrix}$ |

Thus, the phase transition in a system is described by a reducible representation $\Xi$, which may be represented as a direct sum of irreducible representations $A_u$ and $D^{*\vec{q}_0}$, that is $\Xi = A_u \oplus D^{*\vec{q}_0}$.

The free energy of the system is obtained from invariance of representation $\Xi = A_u \oplus D^{*\vec{q}_0}$. According to the condition of invariance, the free energy functional is expressed by

$$\Phi = \frac{1}{d}\int_0^d f(x)dx, \tag{1}$$

$$f(x) = \frac{\alpha}{2}\rho^2 + \frac{\beta}{4}\rho^4 + \gamma\rho^n \cos n\varphi - \delta_1\rho^2(\frac{d\varphi}{dx}) + \frac{k}{2}\rho^2(\frac{d\varphi}{dx})^2 +$$
$$+ 2\xi\rho^{n/2} P\cos(n/2)\varphi + \frac{aP^2}{2} + \frac{bP^4}{4} + \frac{\delta}{2}\left(\frac{dP}{dx}\right)^2 + \frac{\lambda}{2}\left(\frac{d^2P}{dx^2}\right)^2 + \frac{v}{2}P^2\left(\frac{dP}{dx}\right)^2 - PE. \tag{2}$$

Here $\alpha = \alpha_0(T - T_0)$, $a = a_0(T - T_0^*)$, $\rho$ and $\varphi$ — amplitude and phase of the order parameter, respectively. $P$ is the polarization along the $b$ axis, $E$ is the external electrical field, $d$ is the period of free energy density $f(x)$. Results of experimental researches of influence of external fields on critical temperature of phase transition in crystals TlGaSe$_2$ and TlInS$_2$ allows to assume that $T_0 = T_0^*$. On seen it is connected with strong interaction of the order parameter — optical soft mode with polarization [10, 13, 14].

The equilibrium phases of the system are determined from the condition of minimum free energy. Thus, for the initial high-symmetry phase with the space group of symmetry $C_{2h}^6$ we obtain that $\rho = P = 0$ and $\varphi$ are undetermined. In high-symmetry phase the temperature dependence of dielectric constant is

$$\chi = C(T - T_i)^{-1}.$$

In the commensurate phase we have $\frac{d\rho}{dx} = \frac{d\varphi}{dx} = \frac{dP}{dx} = 0$ and from the minimization condition we obtain:

1. $\rho = 0, P \neq 0, \Rightarrow C_2^3, N = 1$

2. $\rho \neq 0, P \neq 0, \sin 4\varphi = 0 \Rightarrow C_2^3, N = 4$

3. $\rho \neq 0, P \neq 0, \cos 4\varphi = 0 \Rightarrow S_2^1, N = 4$

4. $\rho \neq 0, P \neq 0, \cos 4\varphi \neq 0 \Rightarrow C_1^1, N = 4$

Here $N = \frac{V_c}{V}$, $V_c$ and $V$ are volumes of an elementary cell in commensurate and high-symmetry phases, respectively.

The temperature dependence for the dielectric constant in commensurate phase is

$$\chi = C(2(T - T_c))^{-1}.$$

Minimization of the functional free energy with respect to $\varphi$ and $P$, leads to following Euler equations:

$$k\frac{d^2\varphi}{dx^2} + n\gamma\rho^{n-2}\sin n\varphi + n\xi P\rho^{\frac{n}{2}-2}\sin\frac{n}{2}\varphi = 0, \tag{3}$$

$$\lambda\frac{d^4P}{dx^4} - (\delta + vP^2)\frac{d^2P}{dx^2} - vP(\frac{dP}{dx})^2 + aP + bP^3 + 2\xi\rho^{\frac{n}{2}}\cos\frac{n}{2}\varphi - E = 0. \tag{4}$$

These last equations may have periodic solutions with different periods $d$. The equilibrium period $d$ is determined by the condition $\delta F / \delta d = 0$, which takes the form

$$\lambda \frac{dP}{dx}\frac{d^3P}{dx^3} - \lambda(\frac{d^2P}{dx^2})^2 - \frac{1}{2}(\delta + vP^2)(\frac{dP}{dx})^2 - \frac{k}{2}\rho^2(\frac{d\varphi}{dx})^2 + \frac{\alpha}{2}\rho^2 + \frac{\beta}{4}\rho^4 + \gamma\rho^n \cos n\varphi +$$
$$+ \frac{a}{2}P^2 + \frac{b}{4}P^4 + 2\xi\rho^{\frac{n}{2}} P \cos\frac{n}{2}\varphi - PE = \Phi \tag{5}$$

Let us notice, that the left part of the equation (5) does not depend on $x$, and represents an integral of energy. The equations (4) and (5) can be used together with equations (3) and (4) in finding the functions $\varphi(x)$ and $P(x)$.

By making the substitution $z(P) = (\frac{dP}{dx})^2$ and using (4), the equation (5) can be written in the following form:

$$\frac{\lambda}{2} z(P)\frac{d^2 z(P)}{dP^2} - \frac{\lambda}{8}(\frac{dz(P)}{dP})^2 - \frac{1}{2}(\delta + vP^2)z(P) + \frac{a}{2}P^2 + \frac{b}{4}P^4 - PE + \frac{\gamma}{2\xi^2}g^2(P) +$$
$$+ Pg(P) - k\rho^2(\frac{1}{4}n^2\rho^n\xi^2 - g^2(P))^{-1} s^2(P) = \Phi_0. \tag{6}$$

Here

$$\Phi_0 = \Phi - \frac{\alpha}{2}\rho^2 - \frac{\beta}{4}\rho^4 - \gamma\rho^8$$

$$g(P) = E - aP - bP^3 + vPz(P) + \frac{1}{2}(\delta + vP^2)\frac{dz(P)}{dP} - \frac{\lambda}{2}[\frac{1}{2}\frac{dz(P)}{dP}\frac{d^2z(P)}{dP^2} + z(P)\frac{d^3z(P)}{dP^3}] \tag{7}$$

$$s(P) = [a + 3bP^2 - vz(P) - 2vP\frac{dz(P)}{dP} - \frac{1}{2}(\delta + vP^2)\frac{d^2z(P)}{dP^2}$$
$$+ \frac{\lambda}{4}[3\frac{dz(P)}{dP}\frac{d^3z(P)}{dP^3} + (\frac{d^2z(P)}{dP^2})^2 + 2z(P)\frac{d^4z(P)}{dP^4}]](z(P))^{\frac{1}{2}}$$

We find the solution in the form

$$Z(P) = \frac{a_0}{4}P^4 + \frac{b_0}{3}P^3 + \frac{c_0}{2}P^2 + d_0 P + e_0. \tag{8}$$

To determine $a_0, b_0, c_0, d_0$ and $e_0$, we substitute (8) into (6) and obtain the following expressions:

$$a_0 = \frac{v}{2\lambda}, \quad b_0 = 0, \quad c_0 = \frac{\delta v - 4b\lambda}{\lambda v}, \quad d_0 = 0, \quad e_0 = \frac{-8a\lambda v^2 + 4\alpha\lambda v^2 + \delta^2 v^2 - 16b^2\lambda^2}{2\lambda v^3},$$

$$\Phi_0 = \frac{16b^3\lambda^2 + 8ab\lambda v^2 - 4b\alpha\lambda v^2 - b\delta^2 v^2}{v^4} \tag{9}$$

If a solution of equation (6) is known, the function $P = P(x)$ can be found from the relation $z(P) = (\frac{dP}{dx})^2$ [2] and therefore we get

$$x = \int [Z(P)]^{-1/2} dP \tag{10}$$

After integration we find:

$$P(x) = \eta sn(px,\kappa), \text{ where } \eta^2 = -\frac{2c_0}{a_0}\frac{\kappa^2}{1+\kappa^2}, \quad p^2 = -\frac{c_0}{2}\frac{1}{1+\kappa^2}, \quad \kappa^2 = \frac{1-(1-\frac{4a_0e_0}{c_0^2})^{\frac{1}{2}}}{1+(1-\frac{4a_0e_0}{c_0^2})^{\frac{1}{2}}}. \quad (11)$$

Here $a_0 > 0$ and the equation $Z(P) = 0$ has four real roots, for what, it is particularly necessary that $c_0 < 0$. Equation (10) has a real solution only for $Z(P) > 0$. The interval of values, for which $Z(P) > 0$, must be bounded, since, otherwise, $P$ would reach infinite values.

Thus, the period of free energy density (1) is determined as follows:

$$d = \frac{4}{p}K(\kappa),$$

where $\kappa \in [0,1]$, and $K(\kappa)$ is the complete elliptic integrals of the first kinds.

Condition (5) can take a different form if it is integrated over $x$ from zero to $d$ and the expression for $\Phi$ is substituted from (1):

$$\int_0^d [2\lambda(\frac{d^2P}{dx^2})^2 + k\rho^2(\frac{d\varphi}{dx})^2 + (\delta + vP^2)(\frac{dP}{dx})^2]dx = 0. \quad (12)$$

The expression (10) defines a temperature dependence of parameter $\kappa$ in an incommensurate phase.

The temperature dependence of dielectric constant $\chi = (\frac{\partial^2\Phi}{\partial E^2})_{\lim E \to 0}$ in the incommensurate phase is expressed as

$$\chi(T) = \frac{\gamma}{2\xi^2} + \frac{2k\rho^2 p}{K(\kappa)}\int_0^d \frac{s^2(P)(n^2\xi^2\rho^n + 4w^2(P))}{(n^2\xi^2\rho^n - 4w^2(P))^3}dx, \quad (13)$$

where $w(P) = g(E=0,P)$. As it follows from the expression (11), in the point of the transition $\chi(T_i) = \frac{\gamma}{2\xi^2}$ and in an incommensurate phase the dielectric constant $\chi = \chi(T)$ weakly depends on temperature.

**Summary**

A large number of works is devoted to experimental studying of temperature dependence of the dielectric constant in crystals TlGaSe$_2$ and TlInS$_2$ [13,14,15]. As well as it was necessary to expect, the results, given in these works, in essence do not differ. The characteristic experimental temperature dependence of the dielectric constant for the TlGaSe$_2$ and TlInS$_2$ crystals is taken from [13] shown in Fig.1 and Fig 2.

Comparisons of the obtained in present research the temperature dependence of dielectric constant with experiment in high-symmetry and commensurate phases show, that the value of the Curie-Weiss constant is $C \approx 2 \times 10^3$ K. That is characteristic for phase transitions with mixed mechanism of phase transitions.

Describing the experimental temperature dependence of dielectric constant in a incommensurate phase by the formula (13), we obtain $\frac{\gamma}{2\xi^2} \approx 100$ and $\frac{a_0\alpha_0 k}{\beta} \approx 1.35$. At comparison of theoretical results with experiment it is necessary to mean that $\varepsilon = 1 + 4\pi\chi$. Besides, the experimental researches show, that $\chi = \chi(T)$ weakly depends on temperature in this phase. Thus, as opposed to transitions of a type II and I in systems with competing interaction in an incommensurate phase dielectric constant weakly depends on temperature.

In a commensurate phase the lattice quadrupled, and space group of symmetry of structure is $C_2^3$. Besides, the commensurate phase is polar.

Thus, the comparison of theoretical results with experimental allows to describe the basic features of phase transitions in crystals TlGaSe$_2$ and TlInS$_2$.

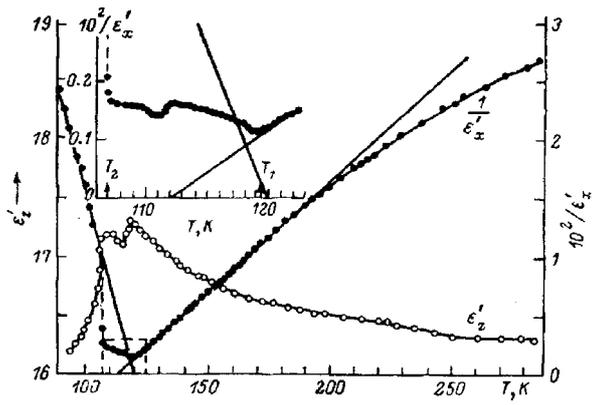

**Fig. 1** Temperature dependences $(\varepsilon'_x)^{-1}$ and $\varepsilon'_z$ of the crystal TlGaSe$_2$ on frequency 100 kHz [13]

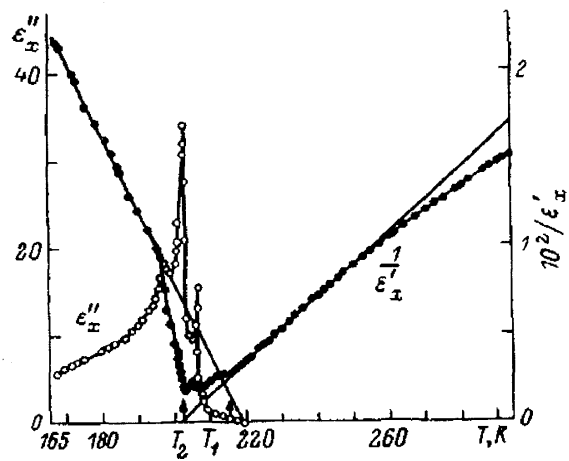

**Fig. 2** Temperature dependences $(\varepsilon'_x)^{-1}$ and $\varepsilon''_x$ of the crystal TlInS$_2$ on frequency 100 kHz [13]